\documentclass{nic-series}

\def\lsim{\raise0.3ex\hbox{$<$\kern-0.75em\raise-1.1ex\hbox{$\sim$}}}
\def\gsim{\raise0.3ex\hbox{$>$\kern-0.75em\raise-1.1ex\hbox{$\sim$}}}

\begin{document} 

\title{QCD phase transition in the chiral limit}

\author{Olaf Kaczmarek, Frithjof Karsch, Anirban Lahiri, \\ Lukas Mazur and Christian Schmidt}

\institute{Fakult\"at f\"ur Physik, Universit\"at Bielefeld, D-33501 Bielefeld, Germany.}

\maketitle

\begin{abstracts}
We present a lattice QCD based determination of the chiral phase transition
temperature in QCD with two massless (up and down) and one strange quark
having its physical mass.  We propose and calculate two novel estimators for the chiral transition
temperature for several values of the light quark masses, corresponding to Goldstone
pion masses in the range of $58~{\rm MeV}\lsim m_\pi\lsim 163~{\rm MeV}$. The chiral
phase transition temperature is determined by extrapolating to vanishing pion
mass using universal scaling analysis. After thermodynamic, continuum and chiral
extrapolations we find the chiral phase transition temperature
$T_c^0=132^{+3}_{-6}$~MeV. We also present some preliminary calculations on interplay
of effective $U_A(1)$ restoration and chiral phase transition towards chiral limit.
\end{abstracts}

\section{Introduction}

For physical values of the two light (up and down) and one heavier (strange)
quark masses strongly interacting matter undergoes a crossover from a low
temperature hadronic regime to a high temperature region that is best described by
quark and gluon degrees of freedom. This smooth crossover between the two asymptotic
regimes is not a phase transition~\cite{Ding:2015ona}. It is characterized by a
pseudo-critical temperature, $T_{pc}$, that has been determined in several numerical
studies of Quantum Chromodynamics (QCD)~\cite{Aoki:2009sc,Bazavov:2011nk,Bonati:2015bha}.
Recent $T_{pc}$ has been calculated precisely from the maximal fluctuations of
several chiral observables gave $T_{pc}= (156.5\pm 1.5)$~MeV~\cite{Bazavov:2018mes}.
On the contrary to the calculation for physical masses, till date one of the outstanding
challenges in QCD thermodynamics is to clarify
the nature of the QCD phase transition that exists in the chiral limit,
i.e. in the limit of vanishing light quark masses, 
$(m_u,m_d) \rightarrow (0,0)$. 
While it is widely expected that the chiral phase transition at vanishing 
values of the  two light quark masses is a second-order transition, belonging 
to the $O(4)$ universality class, a subtle role  is played by 
the $U(1)_A$ axial anomaly \cite{Pisarski:1983ms}.
If the $U(1)_A$ symmetry, which is broken
in the QCD vacuum, does not get \textquotedblleft effectively restored\textquotedblright\ at high 
temperature, the transition indeed will be in the universality of $3$-$d$,
$O(4)$ spin models. However, if the $U(1)_A$ symmetry breaking effects are
small already at the chiral phase transition, at which the chiral condensate
vanishes, the phase transition may turn out to be first order 
\cite{Pisarski:1983ms}, although a second-order transition belonging to larger 3-$d$ universality
class~\cite{Butti:2003nu,Grahl:2013pba,Pelissetto:2013hqa,Sato:2014axa} may become of relevance.
If the chiral phase transition is first
order then a second order phase transition, belonging to the 3-$d$ $Z(2)$
universality class, would occur for $m_l^c>0$. When decreasing the light to strange
quark mass ratio, $H=m_l/m_s$, towards zero, this would give rise to diverging susceptibilities
already for some critical mass ratio $H_c=m_l^c/m_s >0$.

\section{Determination of chiral phase transition temperature}
\subsection{Observables}
In QCD chiral symmetry is spontaneously broken at low temperatures
The corresponding order parameter named chiral condensate is obtained as the derivative of the
partition function, $Z(T,V,m_u,m_d,m_s)$, with respect to the quark 
mass $m_f$ of flavor $f$,
\begin{equation}
\langle \bar\psi \psi\rangle_f = \frac{T}{V} \frac{\partial 
\ln Z(T,V,m_u,m_d,m_s)}{\partial m_f} \; .
\label{pbp}
\end{equation}
In the chiral limit, $m_l\rightarrow 0$, the light quark chiral condensate, 
$\langle \bar\psi \psi\rangle_l= (\langle \bar\psi \psi\rangle_u+\langle \bar\psi \psi\rangle_d)/2$, 
is an exact order parameter for the chiral phase transition.
We take care of additive and multiplicative renormalization by introducing \cite{Bazavov:2011nk} a
combination made out of the light and strange quark chiral condensates,
\begin{equation}
M = 2 \left( m_s \langle \bar\psi \psi\rangle_l - m_l \langle \bar\psi \psi\rangle_s
\right)/f_K^4 \; ,
\label{M}
\end{equation}
where 
$f_K=156.1(9)/\sqrt{2}$~MeV,
is the kaon decay constant, 
which we use as normalization constant to define a
dimensionless order parameter $M$. Derivative of $M$ with 
respect to the light quark masses defines the 
renormalized chiral susceptibility, 
\begin{eqnarray}
	\hspace*{-0.2cm}\chi_M &=& \left. 
	m_s (\partial_{m_u}+\partial_{m_d}) M \right|_{m_u=m_d}
\nonumber \\
&=&m_s( m_s \chi_l - 2 \langle \bar\psi \psi\rangle_s -4 m_l \chi_{su}
)/f_K^4\; ,
\label{chim}
\end{eqnarray}
with
$\chi_{fg}=\partial_{m_f}  \langle \bar\psi \psi\rangle_g$
and $\chi_l= 2(\chi_{uu}+\chi_{ud})$.
$M$ and $\chi_M$ are renormalization group invariant quantities
when terms proportional to the logarithm of light quark masses
can be neglected.

\begin{figure}[!t]
\centering
\includegraphics[width=0.70\textwidth]{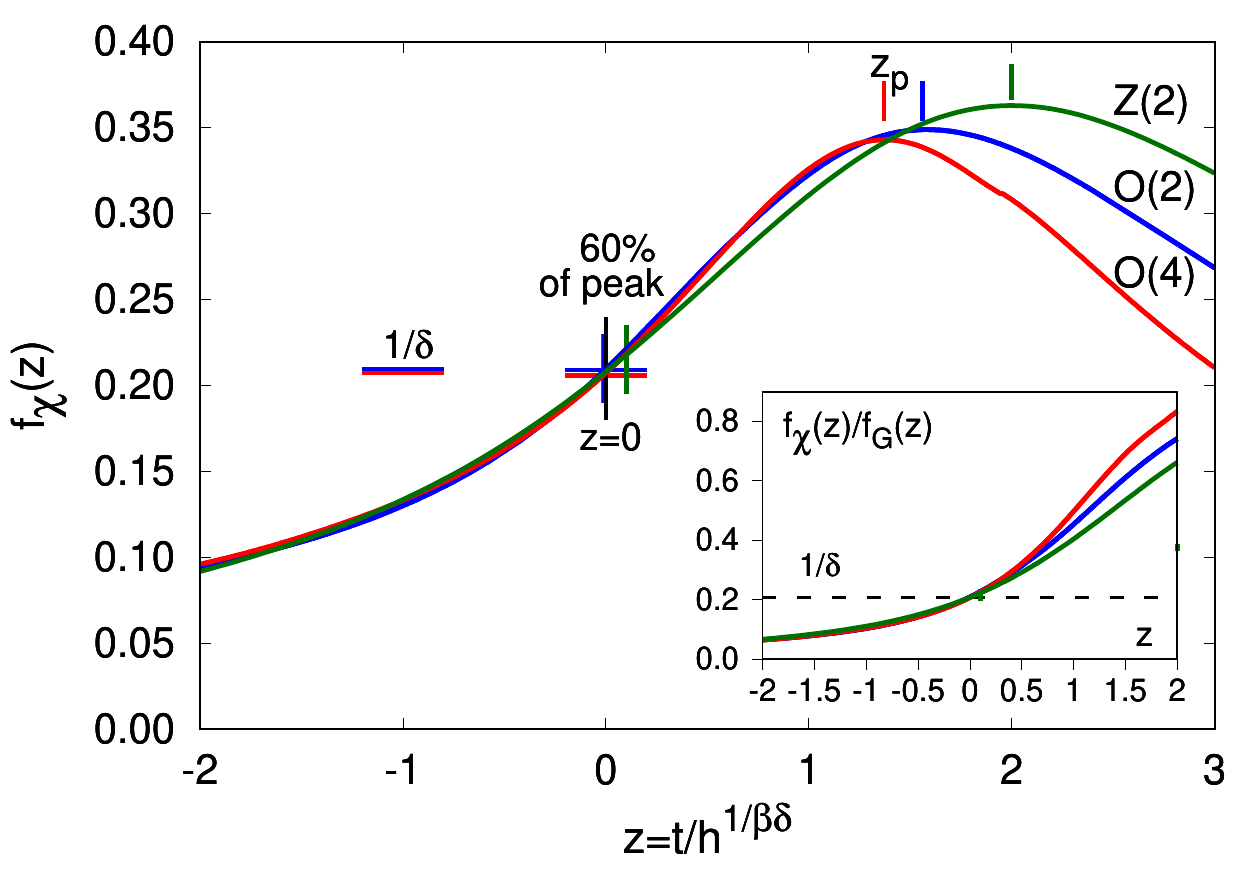}
\caption{Scaling functions for the 3-$d$ $Z(2)$, $O(2)$ and $O(4)$
universality classes. The position $z_p$ of the peak of the scaling functions
(vertical lines) and the position $z_{60}$ where the scaling function attains
60\% of its maximal value (crosses) are shown.
Lines close to $z=-1$ show $1/\delta$ for these three universality classes, 
which agree to better than 1\%. The inset shows the ratio of scaling functions,
$f_\chi(z)/f_G(z)$, used in determinations of the chiral phase transition
temperature.
}
\label{fig:scaling}
\end{figure}

Near a $2^{nd}$ order phase transition $M$ and $\chi_M$ can be described by
universal finite-size scaling functions $f_G(z,z_L)$ and $f_\chi(z,z_L)$
where the scaling variables in arguments are defined as
$z=t/h^{1/\beta\delta}$ and $z_L= l_0/(Lh^{\nu/\beta\delta})$,
with $t=(T-T_c^0)/(t_0 T_c^0)$ being the reduced temperature; $h=H/h_0$
with $H=m_l/m_s$
denotes the symmetry breaking field and $L$ being the linear extent of
the system, $L\equiv V^{1/3}$. The normalization constants $t_0,\ h_0$ and $l_0$ that appear in
definition of the scaling variables are non-universal parameters.
Before approaching the chiral limit, one also needs to take a proper
thermodynamic limit, $V \rightarrow \infty$ of any calculation.

Although close to a critical point at $(z,z_L)=(0,0)$, $M$ and $\chi_M$
can be described by the universal scaling functions, but away from the
critical point they also receive contributions from corrections-to-scaling
\cite{Hasenbusch:2000ph,Engels:2000xw} and regular terms.
With this we may write
\begin{eqnarray}
M &=& h^{1/\delta} f_G(z,z_L) + f_{sub}(T,H,L) \; , \nonumber \\
\chi_M &=& h_0^{-1} h^{1/\delta-1} f_\chi(z,z_L) +\tilde{f}_{sub}(T,H,L) \; .
\label{scale}
\end{eqnarray}
where $f_{sub}(T,H,L)$ and $\tilde{f}_{sub}(T,H,L)$ denotes the above-mentioned
sub-leading contributions for $M$ and $\chi_M$ respectively.

A commonly defined estimator for pseudo-critical temperature corresponds to
the peak in the scaling function $f_\chi(z,z_L)$ for large enough system sizes.
Within the scaling regime the peak is located at $z=z_p(z_L)$, which defines $T_p$
\cite{Bazavov:2011nk},
\begin{equation}
T_p(H,L)=
T_c^{0} \left( 1+ \frac{z_p(z_L)}{z_0} H^{1/\beta\delta} \right)\ +\ \text{sub-leading} \; ,
\label{Tpc}
\end{equation}
with $z_0=h_0^{1/\beta\delta}/t_0$.
The first term accounts for the universal quark mass dependence of $T_p$. Apart from that
contributions from corrections-to-scaling and regular terms will be there, shifting
the peak-location of the chiral susceptibilities. These contributions together
has been denoted as \textquoteleft sub-leading\textquoteright\ in Eq.~\ref{Tpc}.

Depending on the magnitude of $z_p/z_0\equiv z_p(0)/z_0$, $T_p(H,L)$
may change significantly with $H$, when approaching towards chiral limit \cite{Bazavov:2011nk}.
So the contribution from the sub-leading terms will be non-negligible which
in turn makes the chiral extrapolation non-trivial. Moreover one has to
deal with more non-universal parameters. Thus it will be really advantageous
to determine $T_c^{0}$ from scaling of such estimators which are defined
with $z\simeq0$. Then by construction $T_p(H,L)$ will have a way milder $H$
dependence which results in rather easier to control $H\rightarrow 0$ extrapolation
to calculate $T_c^0$.

Here we consider two such estimators \cite{Ding:2019prx,Ding:2018auz} for $T_c^0$, defined close to or at $z=0$,
in the thermodynamic limit.  We define temperatures $T_\delta$ and $T_{60}$ through the following conditions:
\begin{eqnarray}
\frac{H \chi_M (T_\delta,H,L)}{M(T_\delta,H,L)} &=&\frac{1}{\delta}  \; ,
\label{ratio}\\
\chi_M(T_{60},H) &=& 0.6 \chi_M^{max} \; .
\label{T60}
\end{eqnarray}
The ratio on the LHS of Eq.~\ref{ratio} has already been introduced in Ref.\cite{Karsch:1994hm} 
as a tool to analyze the chiral transition in QCD.
$T_{60}$ will corresponds to the temperature left to the peak of $\chi_M$; {\it i.e.}\ $T_{60}<T_p$.
These pseudo-critical temperatures $T_X$, defined through Eq.~\ref{ratio} and Eq.~\ref{T60},
are already close to $T_c^0$ for finite $H$ and $L^{-1}$ because they involve $z_X(z_L)$
which either vanishes or stay close to zero in the $L^{-1}\rightarrow 0$ limit.
To be precise, $z_\delta\equiv z_\delta(0)=0$ and $z_{60}\equiv z_{60}(0) \simeq 0$.
Some values for $z_{60}$, along with the corresponding scaling functions in the thermodynamic
limit have been shown in Fig.~\ref{fig:scaling} for relevant universality classes.

\begin{figure*}[!t]
\centering
\includegraphics[width=0.70\textwidth]{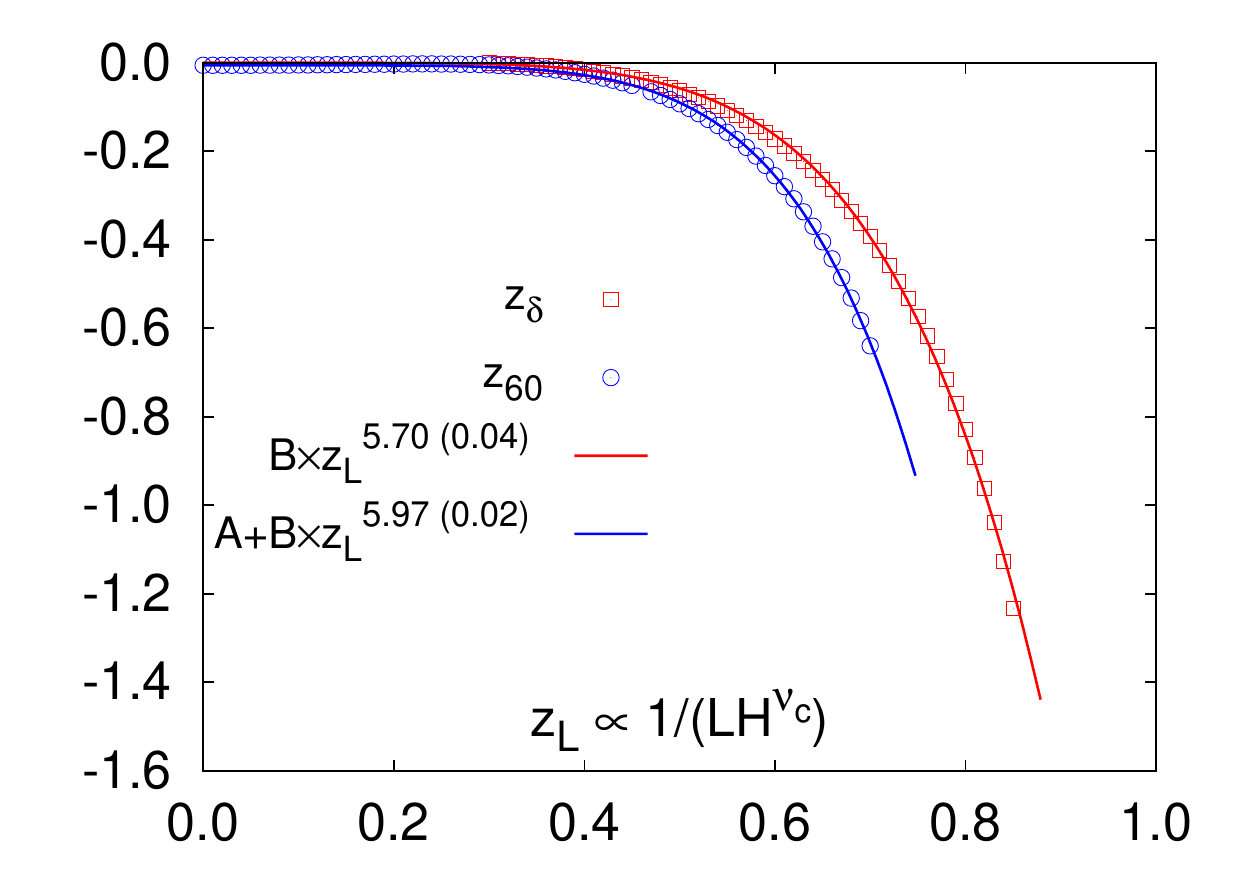}
\caption{Universal functions $z_{\delta} (z_L)$ and $z_{60} (z_L)$ calculated from the finite-size
scaling functions $f_G(z,z_L)$, $f_\chi(z,z_L)$ determined in Ref.~\cite{Engels:2014bra}.
}
\label{fig:zXzL}
\end{figure*}

Ignoring possible contributions from 
corrections-to-scaling, and keeping in $f_{sub}$ only the leading 
$T$-independent, infinite volume regular contribution proportional to $H$,
we then find for the pseudo-critical temperatures \cite{Ding:2019prx},
\begin{eqnarray}
T_X(H,L)= 
T_c^{0} \left( 1+ \left( \frac{z_X(z_L)}{z_0} \right) 
H^{1/\beta\delta} \right)
+c_X H^{1-1/\delta+1/\beta\delta}\;\; ,\;\; X=\delta,\ 60
 \; .
\label{TX}
\end{eqnarray}

The universal functions, $z_X (z_L)$ may directly be determined from the 
ratio of scaling functions, $f_\chi(z_\delta,z_L)/f_G(z_\delta,z_L)=1/\delta$
and $f_\chi(z_{60},z_L)/f_\chi(z_p,z_L)=0.6$, respectively. 
In Fig.~\ref{fig:zXzL} we show the calculation of the universal functions $z_X (z_L)$
along with the optimal parameterized form, for 3-$d$, $O(4)$ universality class using
the finite-size scaling functions $f_G(z,z_L)$, $f_\chi(z,z_L)$ determined in Ref.~\cite{Engels:2014bra}.

We will present here results on $T_\delta$ and $T_{60}$ obtained in lattice 
QCD calculations \cite{Ding:2019prx}.
We calculated the chiral order parameter $M$ and the chiral
susceptibility $\chi_M$ (Eqs.~\ref{M} and \ref{chim})
in $(2+1)$-flavor QCD with degenerate up and down quark masses ($m_u=m_d$).
Our calculations are performed with the Highly Improved Staggered Quark (HISQ) action
\cite{Follana:2006rc} in the fermion sector along with the Symanzik improved gluon action.
The strange quark mass has been tuned to its physical value \cite{Bazavov:2014pvz} and the
light quark mass has been varied in a range $m_l\in [m_s/160:m_s/20]$
corresponding  to Goldstone pion masses in the range $58~{\rm MeV}\lsim m_\pi \lsim 163~{\rm MeV}$.
At each temperature we performed calculations on lattices of size
$N_\sigma^3 N_\tau$ for three different values of the lattice cut-off, 
$aT = 1/N_\tau$, with $N_\tau=6,\ 8$ and $12$. 
The spatial lattice extent, $N_\sigma=L/a$, has been varied in the range 
$4\le N_\sigma / N_\tau \le 8$. For each $N_\tau$ we analyzed the volume dependence
of $M$ and $\chi_M$ in order to perform controlled infinite volume extrapolations.

\subsection{Results}

\begin{figure*}[!t]
\centering
\includegraphics[width=0.70\textwidth]{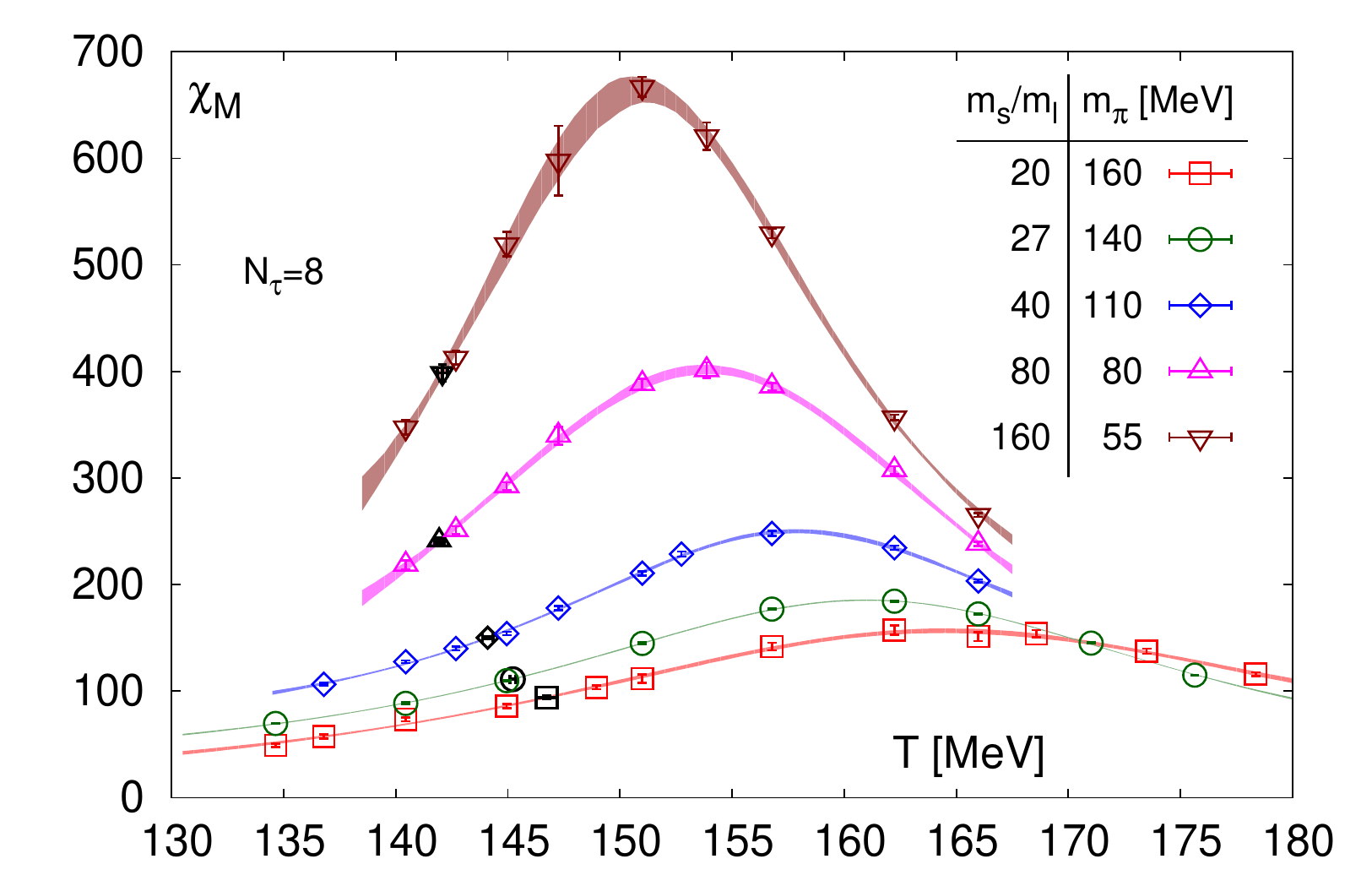}
\caption{Quark mass dependence of
the chiral susceptibility on lattices with temporal extent
$N_\tau=8$ for several values of the light quark masses.
The spatial lattices extent $N_\sigma$ is
increased as the light quark mass decreases: $N_\sigma= 32$~
$(H^{-1}=20,\ 27)$, 40~$(H^{-1}=40)$, 56~$(H^{-1}=80,\ 160)$.
Black symbols mark the points corresponding
to 60\% of the peak height. 
Figure is taken from \cite{Ding:2019prx}.
}
\label{fig:sus_mass}
\end{figure*}

\begin{figure*}[!t]
\centering
\includegraphics[width=0.70\textwidth]{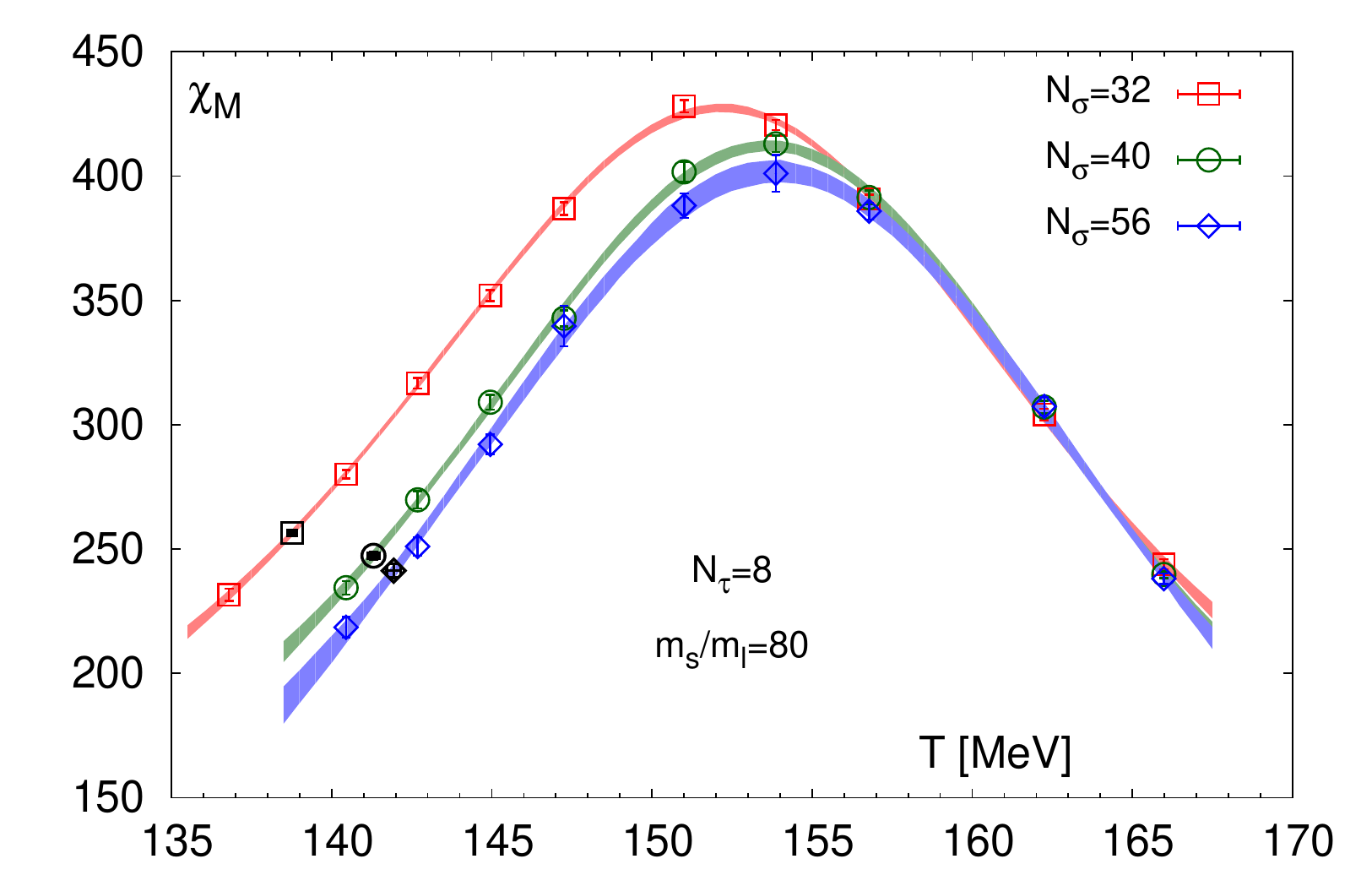}
\caption{Volume dependence of
the chiral susceptibility on lattices with temporal extent
$N_\tau=8$ for three different spatial
lattice sizes at $H=1/80$. Black symbols mark the points corresponding
to 60\% of the peak height. 
Figure is taken from \cite{Ding:2019prx}.
}
\label{fig:sus_vol}
\end{figure*}

We show results for $\chi_M$ in Fig.~\ref{fig:sus_mass}, on lattices with temporal
extent $N_\tau=8$ for $5$ different values of the quark mass ratio, 
$H=m_l/m_s$, and the largest lattice available for each $H$. 
The increase of the peak height, $\chi_M^{max}$, with decreasing $H$ is 
consistent with the expected behavior,
$\chi_M^{max} \sim H^{1/\delta-1}+ const.$, with  $\delta \simeq 4.8$
within rather large uncertainty which restricts a precise
determination of $\delta$.

\begin{figure*}[!t]
\centering
\includegraphics[width=0.70\textwidth]{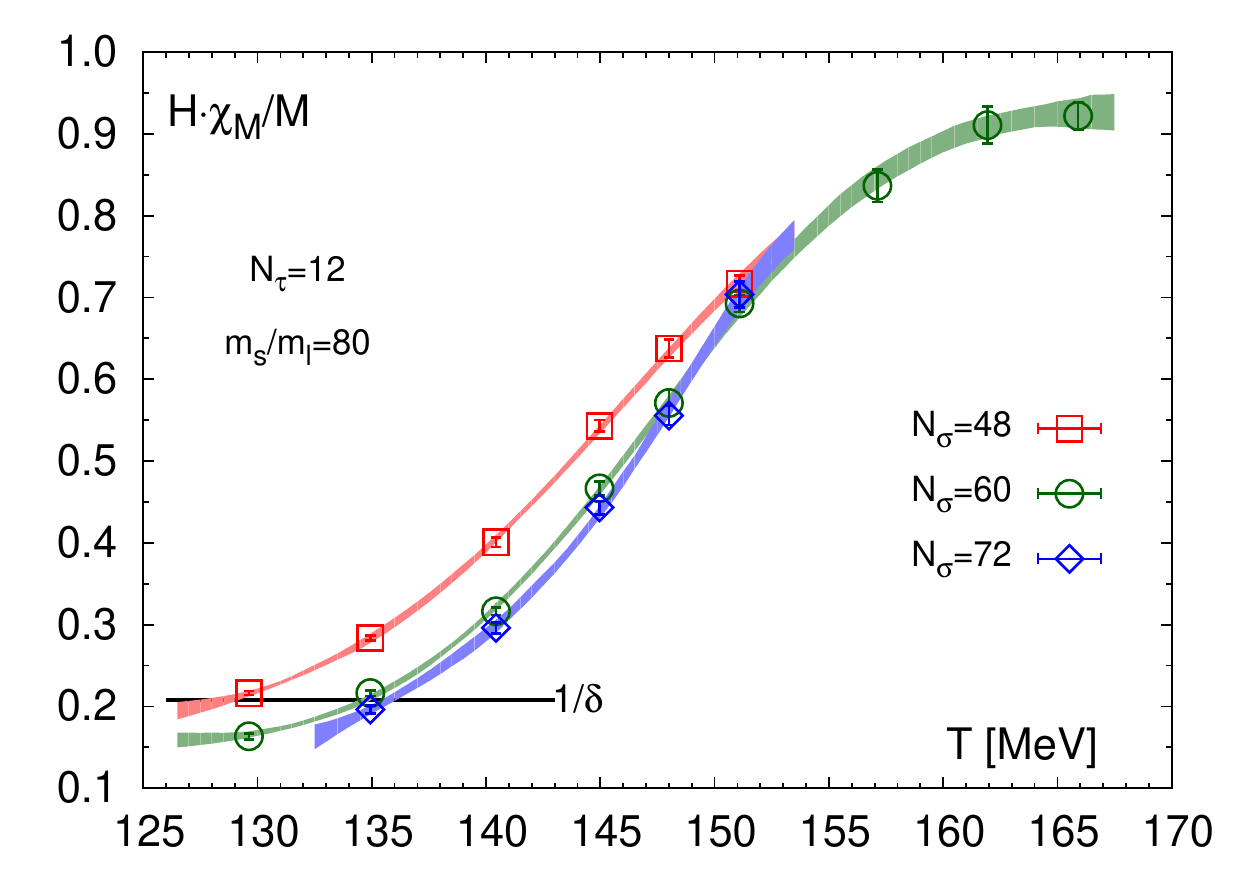}
\caption{
The ratio $H\chi_M/M$ versus temperature for $N_\tau=12$,
$m_l/m_s=1/80$ and different spatial volumes.
Figure is taken from \cite{Ding:2019prx}.
}
\label{fig:HchiM}
\end{figure*}

\begin{figure*}[!t]
\centering
\includegraphics[width=0.70\textwidth]{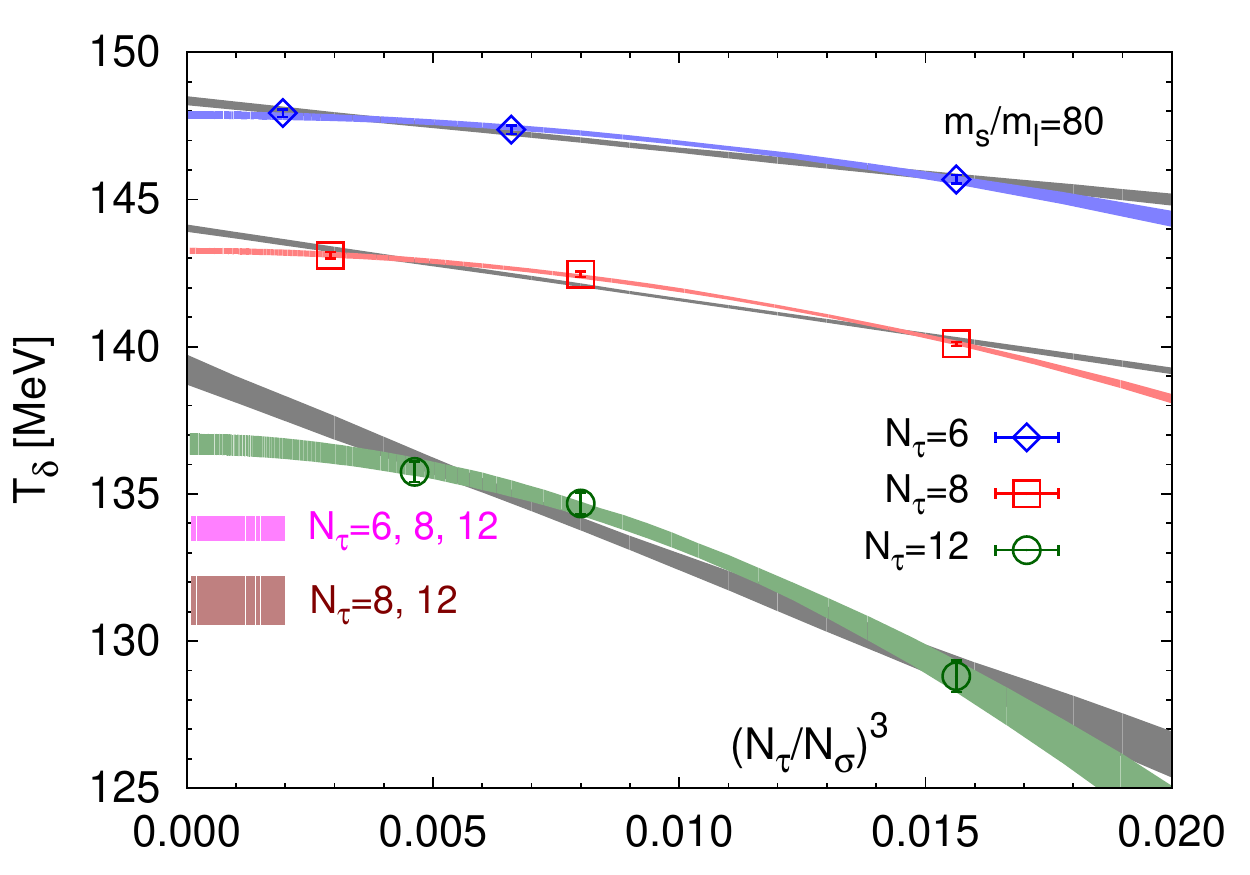}
\caption{
Infinite volume extrapolations based on an $O(4)$ finite-size scaling
ansatz (colored bands) and fits linear in $1/V$ (grey bands). Horizontal
bars show the continuum extrapolated results for $H=1/80$.
Figure is taken from \cite{Ding:2019prx}.
}
\label{fig:Tdeltavolm80}
\end{figure*}

\begin{figure*}[!t]
\centering
\includegraphics[width=0.70\textwidth]{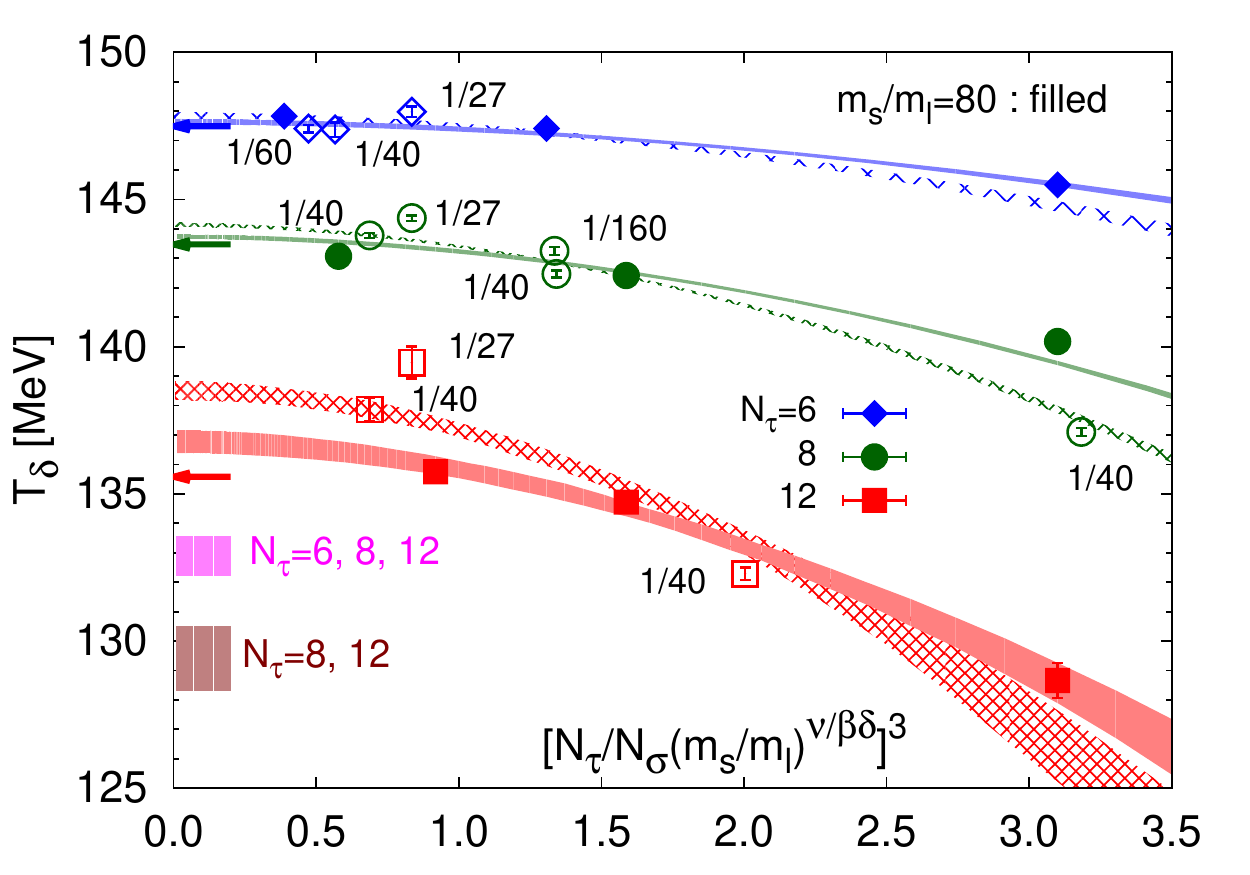}
\caption{
Finite size scaling fits for $T_\delta$ based on all
data for $H\le 1/27$ and all available volumes. Arrows show chiral limit
results at fixed $N_\tau$ and horizontal
bars show the continuum extrapolated results for $H=0$.
Figure is taken from \cite{Ding:2019prx}.
}
\label{fig:Tdeltajoint}
\end{figure*}

In Fig.~\ref{fig:sus_vol} we show the volume dependence of $\chi_M$ for 
$H=1/80$ on lattices with temporal extent $N_\tau=8$ and for $N_\sigma/N_\tau=4,\ 5$ and $7$. 
Similar results have also been obtained for $N_\tau=6$ and $12$.
It is important to note that $\chi_M^{max}$ decreases slightly with increasing volume,
contrary to what one would expect to find at or close to a $1^{st}$
order phase transition. In fact this trend seems to be consistent with the
behavior seen for $O(4)$ universality class finite-size scaling functions \cite{Engels:2014bra}
Our current results, thus, suggest a continuous phase transition at $H_c=0$.

\begin{figure}[h]
\centering
\includegraphics[width=0.70\textwidth]{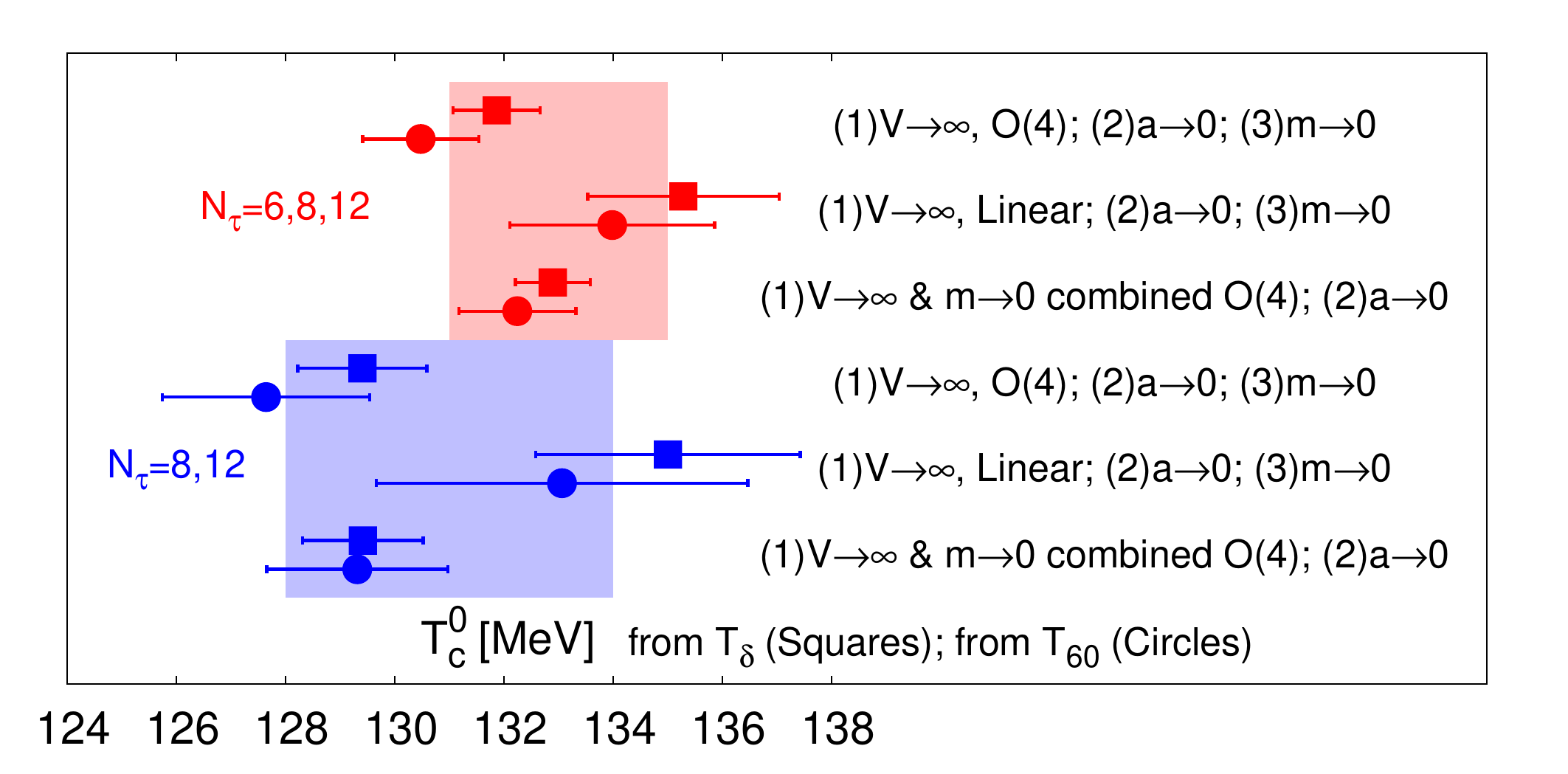}
\caption{Summary of fit results. The order of different limits taken,
described in the main text, is written beside each pair of closest points.
For such a pair squares stands for calculations using $T_{\delta}$ and
circles stands for calculations using $T_{60}$.
Upper half of the points corresponds to the calculations when
continuum extrapolation uses all three $N_{\tau}$ and presented
in red. Similarly lower half, presented in blue, corresponds to
the calculations where only two largest $N_{\tau}$ have been used
in continuum extrapolation.
Figure is taken from \cite{Ding:2019prx}.
}
\label{fig:final}
\end{figure}

Using results for $\chi_M$ and $M$ we constructed the
ratios $H\chi_M /M$ for different lattice sizes and several values of the
quark masses. In Fig.~\ref{fig:HchiM} this ratio has been shown
on the $N_\tau=12$ lattices with $H=1/80$ which is the lowest mass for for this $N_\tau$.
The intercepts with the horizontal line at $1/\delta$ define $T_\delta(H,L)$. 
For $H=1/80$ and each of the $N_\tau$,
we have results for three different volumes on which we can extrapolate
$T_\delta(H,L)$ to the thermodynamic limit. We performed such extrapolations
using (i) the $O(4)$ ansatz given in Eq.~\ref{TX} which is appropriate when the
singular part dominates the partition function and (ii) an extrapolation in $1/V$
which is appropriate if, for large $L$, the volume dependence predominantly arises from regular terms.
In the former case we use the approximation $z_\delta(z_L) \sim z_L^{5.7}$, as shown in Fig.~\ref{fig:zXzL}.
This parameterizes well the finite-size dependence of $T_\delta$ in the scaling regime.
The resulting volume extrapolations are shown in Fig.~\ref{fig:Tdeltavolm80}.
For fixed $H$ the results tend to approach the infinite volume limit more rapidly 
than $1/V$, which is in agreement with the behavior expected from the ratio of
finite-size scaling functions. 
The resulting continuum limit extrapolations in $1/N_\tau^2$ based on
data for with and without $N_\tau=6$ 
are shown as horizontal bars in this figure. 
A similar analysis is performed for $H=1/40$.
Finally, we extrapolate the 
continuum results for $T_\delta(H,\infty)$ with $H=1/40$
and $1/80$ to the chiral limit using Eq.~\ref{TX}
with $z_\delta(0)=0$.
Results obtained from these extrapolation chains, which involve
either an $O(4)$ or $1/V$ ansatz for the infinite volume extrapolation,
and continuum limit extrapolations performed using all three $N_\tau$
or the two largest $N_\tau$, lead to chiral transition temperatures $T_c^0$ 
in the range ($128$-$135$)~MeV.
A complete summary of the resulting values for $T_c^0$ are depicted in Fig.~\ref{fig:final}.

Since the fits in Fig.~\ref{fig:Tdeltavolm80} suggest that the
$O(4)$ scaling ansatz is appropriate to account for the finite volume
effects already at finite $N_\tau$, we can attempt 
a joint infinite volume and chiral extrapolation of all data available
for different light quark masses and volumes at fixed $N_\tau$.
This utilizes the quark mass dependence of finite-size corrections,
expressed in terms of $z_L$. The main difference of this method
from the earlier one is here chiral extrapolation is done before
taking the continuum limit. Using the scaling ansatz given in
Eq.~\ref{TX}, it also allows to account 
for the contribution of a regular term in a single fit. 
Fits for fixed $N_\tau$ based on this ansatz, using data for all available 
lattice sizes and $H\le 1/27$, are shown in Fig.~\ref{fig:Tdeltajoint}.
Bands for $H=1/40$ and $1/80$ are shown in
the figure. As can be seen, for $H=1/80$, these bands compare well 
with the fits shown in Fig.~\ref{fig:Tdeltavolm80}.
For each $N_\tau$ an arrow shows
the corresponding chiral-limit result, $T_\delta(0,\infty)$. 
We extrapolated these chiral-limit
results to the continuum limit and estimated systematic errors again by
including or leaving out data for $N_\tau=6$. 
The resulting $T_c^0$, shown in Fig.~\ref{fig:final}, are in complete
agreement with the corresponding numbers obtained by first taking the
continuum limit and then taking the chiral limit.
Within the current accuracy these two limits are interchangeable. 

Similarly we analyzed results for $T_{60}$ on all data sets
following the same strategy as for $T_\delta$. As can be 
seen in Fig.~\ref{fig:final}, we find for each extrapolation
ansatz that the resulting values for $T_c^0$ 
agree to better than 1\% accuracy with the corresponding values 
from $T_\delta$.
This suggests that the chiral susceptibilities used for this analysis
reflect basic features of the $O(4)$ scaling functions. 

Performing continuum extrapolations by either including or discarding
results obtained on the coarsest ($N_\tau=6$) lattices leads to
a systematic shift of about ($2$-$3$)~MeV in the estimates for $T_c^0$.
This is reflected in the displacement of the two colored bands
in Fig.~\ref{fig:final}, which show averages for $T_c^0$ obtained with our 
different extrapolation ans\"atze.
Averaging separately over results for $T_\delta$ and $T_{60}$ 
obtained with both continuum extrapolation
procedures and including this systematic effect
we find for the chiral phase transition temperature,
\begin{equation}
T_c^0 = 132^{+3}_{-6}~{\rm MeV} \; .
\label{Tcfinal}
\end{equation}

\section{Towards an understanding of anomalous \boldmath$U_A(1)$ symmetry restoration}

\subsection{Observable}
The chiral susceptibility, defined after Eq.~\ref{chim} receives contributions
from a disconnected and connected part which are related to quark-line disconnected 
and connected Feynman diagrams \cite{Bazavov:2011nk}, {\it i.e.}
\begin{equation}
\chi_{l}=\chi_{l,disc}+\chi_{l,conn}
\end{equation}
with
\begin{eqnarray}
\chi_{l,disc} &=&
{{N_f^2} \over 16 N_{\sigma}^3 N_{\tau}} \left\{
\langle\bigl( {\rm Tr} D_l^{-1}\bigr)^2  \rangle -
\langle {\rm Tr} D_l^{-1}\rangle^2 \right\}
\label{chi_dis} \; , \\
\noalign{and}
\chi_{l,conn} &=&
-\frac{N_f}{4N_\sigma^3N_\tau} \langle {\rm Tr} D_l^{-2}\rangle   \; .
\label{chi_con}
\end{eqnarray}
Here $D_l$ denotes the light quark, staggered fermion Dirac matrix.
A crucial role in the analysis of the temperature dependence of the
axial anomaly and its overall strength is played by the
disconnected chiral susceptibility in the following way.

$U(1)_A$ symmetry relates the susceptibilities of the pion and the scalar
iso-triplet delta meson. In a chirally symmetric system one has the degeneracy
between pion and iso-scalar meson whose susceptibility is same as chiral susceptibility.
The disconnected chiral susceptibility, $\chi_{l,disc}$, is directly related 
to $U(1)_A$ in the following way,
\begin{eqnarray}
\chi_\pi - \chi_\delta &=& \chi_{l,disc} + (\chi_\pi - \chi_\sigma ) \; .
\nonumber \\
&=& \chi_{l,disc}, ~~ \text{in chirally restored phase.}
\label{dif}
\end{eqnarray}
which says in a chirally symmetric background $\chi_{l,disc}$ vanishes
when $U(1)_A$ is effectively restored. This is the strategy we are going to take
and here we present some preliminary results in this direction.

\subsection{Results}

\begin{figure}[!t]
\begin{center}
\includegraphics[width=0.70\textwidth]{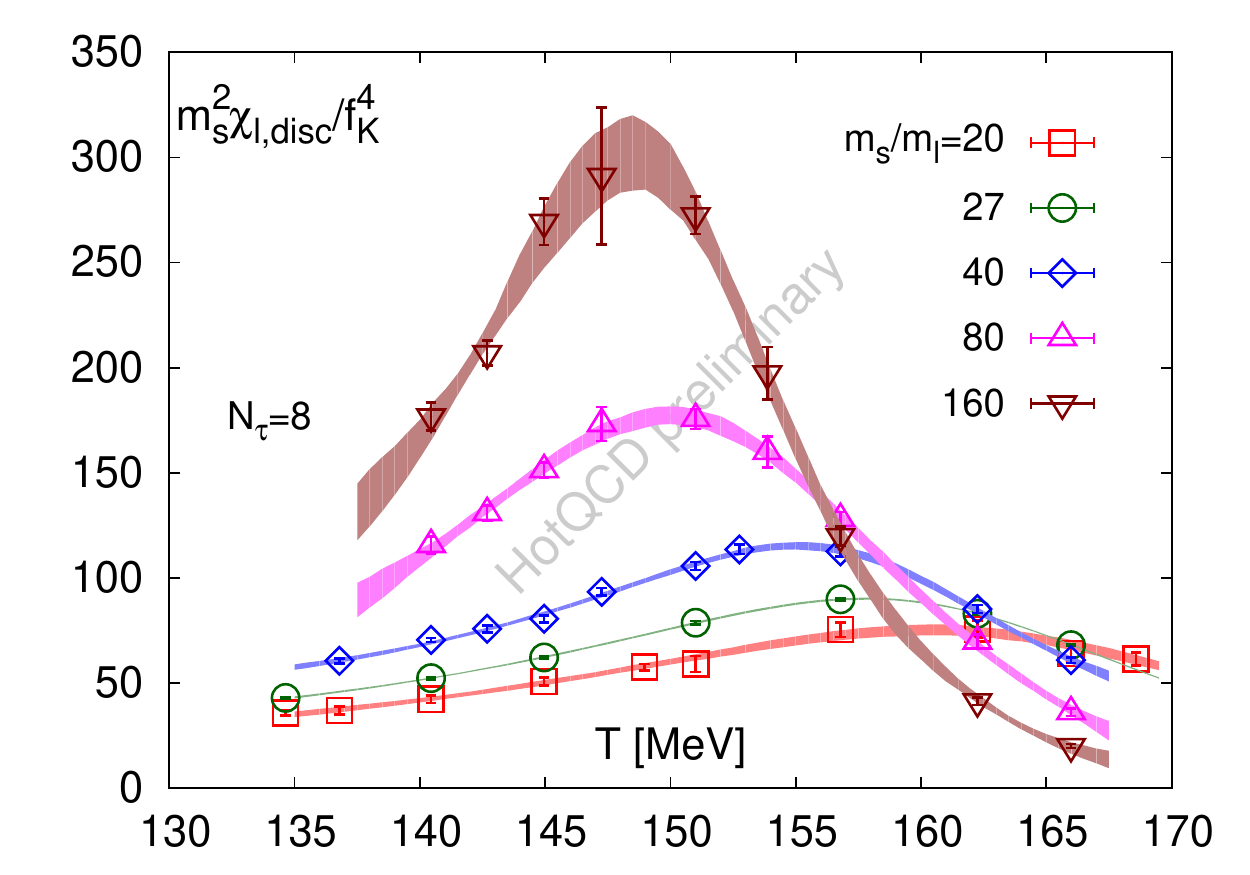}
\end{center}
\caption{The disconnected chiral susceptibility obtained
from calculations in
(2+1)-flavor QCD with the HISQ action on lattices of $N_\tau=8$,
at several values of the light quark mass.
The susceptibility has
been multiplicatively renormalized by multiplying with the square of the
strange quark mass and the kaon decay constant $f_K$ has been used to
set the scale for the susceptibility as well as the temperature.
}
\label{fig:chidiscT}
\end{figure}

\begin{figure}[!t]
\begin{center}
\includegraphics[width=0.70\textwidth]{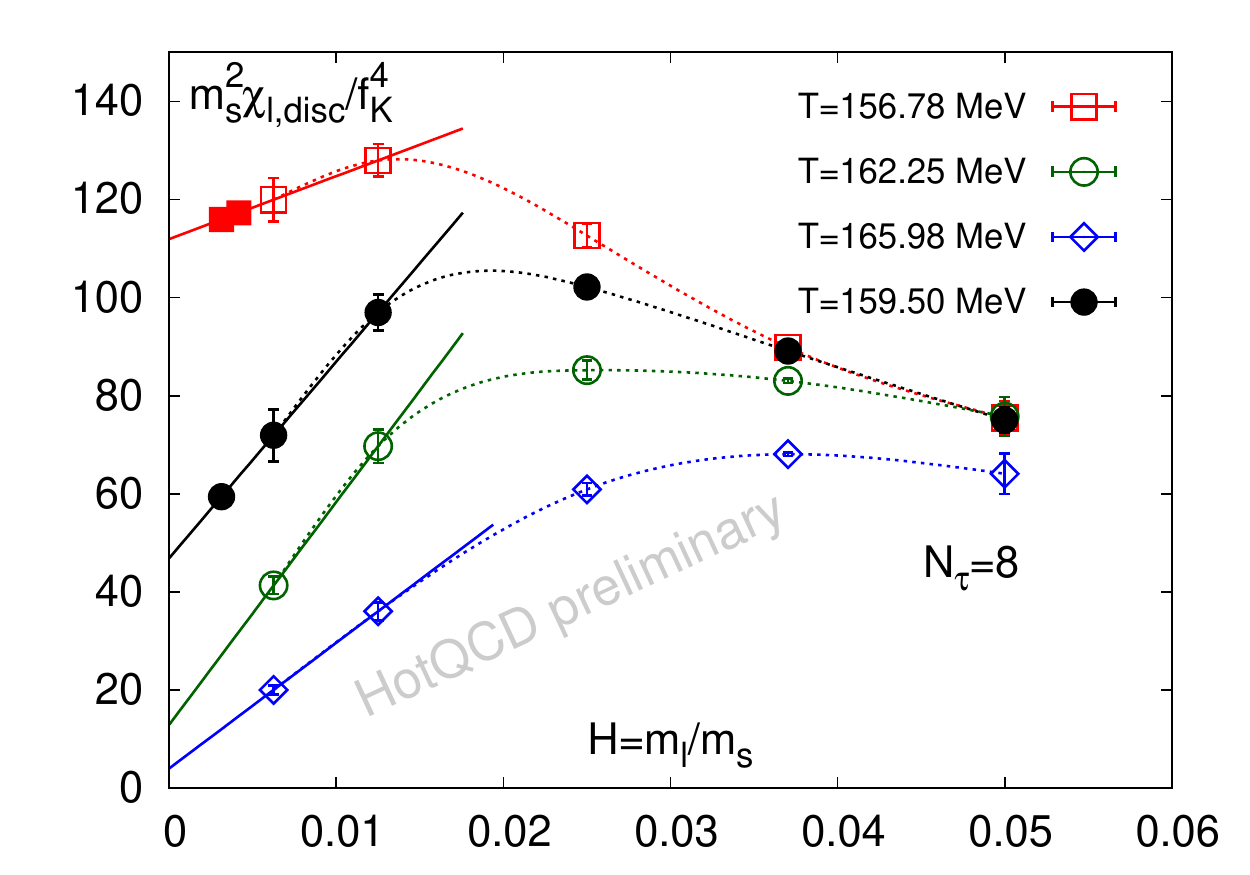}
\end{center}
\caption{
The quark mass dependence of the disconnected chiral
susceptibility at three values of the temperature
obtained on lattices of size $N_\sigma^3\times 8$ with $N_\sigma =
32 - 56$. Shown are results for a light ($m_l$) to strange ($m_s$) 
quark mass ratio $1/160\le m_l/m_s\le 1/20$.
The open symbols are the data points on which direct calculation is
available and solid symbols correspond to those points where interpolation
or extrapolation has been used.
Dotted lines are drawn to guide the eye and the solid lines are drawn
from a linear fit using the points corresponding to $H=1/80$ and $H=1/160$.
}
\label{fig:chidiscm}
\end{figure}

Existing calculations with staggered fermions \cite{Bazavov:2019www,Cheng:2010fe,Ohno:2012br}
as well as overlap and M\"obius domain 
wall \cite{Buchoff:2013nra,Bazavov:2012qja,Tomiya:2016jwr}
and Wilson \cite{Brandt:2016daq} fermions show evidence for effective restoration of anomalous $U_A(1)$
symmetry above the pseudo-critical temperature $T_{pc}$, {\it i.e.}\ at about $(1.2-1.3) T_{pc}$.

At non-zero values of the light quark masses the disconnected chiral 
susceptibility $\chi_{l,disc}(T)$ and the total chiral
susceptibility $\chi_{m}(T)$ share the same qualitative properties.
I.e. $\chi_{dis}(T)$  has a pronounced peak close
to that of the total chiral susceptibility which also rises with
decreasing values of the quark mass. When one approaches the chiral limit
these two peaks start coming more close to each other and in the chiral
limit both total and disconnected chiral susceptibility diverges at $T_c^0$.
In particular, the location of the peak in $\chi_{l,disc}$
has also been used in calculations with chiral fermions (Domain
Wall Fermions) to locate the QCD transition temperature \cite{Bhattacharya:2014ara}.
Also in the chiral limit it generally is expected that $\chi_{l,disc}$ 
remains non-zero in the high temperature
chirally symmetric phase and that it diverges as 
the chiral phase transition temperature is approached from above. 
In Fig.~\ref{fig:chidiscT} we show
the quark mass dependence of the disconnected chiral susceptibility obtained 
from calculations in
(2+1)-flavor QCD with the HISQ action on lattices of $N_\tau=8$,
where the above-mentioned properties are apparent.
While the figure shows in the vicinity of the transition temperature
the expected quark mass dependence, i.e. $\chi_{l,disc}$ rises with 
decreasing quark mass, the figure also shows
that at temperatures $T~\gsim~165$~MeV 
the quark mass dependence in the chiral susceptibility seems to 
be inverted. The disconnected chiral susceptibility becomes smaller
as the quark mass decreases.

In Fig.~\ref{fig:chidiscm} we show preliminary results of $\chi_{l,disc}$
as a function of scaled light quark masses for various temperatures which
are above chiral critical temperature, obtained on lattices with temporal extent $N_\tau=8$
and spatial lattice sizes up to $N_\sigma = 56$. Some of these high temperature
results, in particular checks on their insensitivity to finite volume 
effects have been confirmed \cite{Ding:2019prx}. One can see at $T\geq 162$ MeV
$\chi_{l,disc}$ almost monotonically decreases and the trend suggests that
in chiral limit within uncertainty it will vanish. Although the situation becomes
more involved for $T\leq 160$ MeV when $\chi_{l,disc}$ first increases with
decreasing quark mass and attains a maximum and then sharply decreases.
Preliminary calculation of Fig.~\ref{fig:chidiscm} suggests that for a range of
temperatures below 160 MeV $\chi_{l,disc}$ will still go to zero after proper chiral
extrapolation but for even lower temperatures chiral extrapolation will give
a finite intercept. In other words, from Fig.~\ref{fig:chidiscm}, it seems to
be quite unlikely that $\chi_{l,disc}$ will vanishes in chiral limit down to $T=T_c^0$.
More calculations are needed to confirm this.

\section{Conclusions}
Based on two novel estimators, we have calculated the chiral
phase transition temperature in QCD with two massless light quarks and a
physical strange quark. Eq.~\ref{Tcfinal} lists our thermodynamic-, continuum- and
chiral- extrapolated result for the chiral phase transition temperature,  which is
about $25$~MeV smaller than the pseudo-critical (crossover)  temperature,
$T_{pc}$ for physical values of the light and strange quark masses.
Preliminary calculations of disconnected chiral susceptibility suggests
that it $U_A(1)$ symmetry remains broken at chiral phase transition.

\section*{Acknowledgments}
This work was supported in part by
the Deutsche Forschungsgemeinschaft (DFG) through the grant 315477589-TRR 211,
the grant 05P18PBCA1 of the German Bundesministerium f\"ur Bildung und
Forschung, grant 283286 of the European Union, the National Natural
Science Foundation of China under grant numbers 11775096 and 11535012,
Furthermore, this work was supported through Contract No.~DE-SC0012704 with the
U.S. Department of Energy, through the Scientific Discovery through Advanced
Computing (SciDAC) program funded by the U.S. Department of Energy, Office of
Science, Advanced Scientific Computing Research and Nuclear Physics and
the DOE Office of Nuclear Physics funded BEST topical collaboration, 
and a Early Career Research Award of the Science and Engineering
Research Board of the Government of India.
Numerical calculations have been made possible through PRACE grants
at CSCS, Switzerland, and at the Gauss Centre for Supercomputing and 
NIC-J\"ulich, Germany as well as grants at CINECA, Italy.
These grants provided access to resources on
Piz Daint at CSCS, at JUWELS at NIC as well as on Marconi at CINECA.
Additional calculations have been performed on 
GPU clusters of USQCD, 
at Bielefeld University, the PC$^2$ Paderborn 
University and the Nuclear Science Computing Center at Central China
Normal University, Wuhan, China. Some data sets have also partly been
produced at the TianHe II Supercomputing Center in Guangzhou.

\end{document}